\documentclass{ws-p8-50x6-00}
\begin{document}

\title{Dicke-superradiance in electronic systems coupled to reservoirs}
\author{T. Brandes$^*$, J. Inoue, and A. Shimizu} 
\address{Institute of Physics, University of Tokyo,
3--8--1 Komaba, Tokyo 153-8902, Japan, and CREST, JST\\
$^*$ (new address) 1. Inst. f. Theor. Physik, Universit\"at Hamburg,
Jungiusstr. 9, D-20355 Hamburg, Germany (brandes@physnet.uni-hamburg.de)}


\maketitle

\abstracts{We present theoretical results for superradiance, i.e. the collective coherent decay
of a radiating system, in semiconductor structures. An optically active region can become superradiant
if a strong magnetic field is applied. Pumping of electrons and holes at a rate $T$ through coupling to external
`reservoirs' leads to a novel kind of oscillations with  frequency $\sim \sqrt{T}$.}

\section{Introduction}
Spontaneous emission is one of the most basic concepts of quantum physics that can be
traced back to such early works as that of Albert Einstein
in 1917 \cite{Ein17}.
Mostly discussed
in the context of an atom  coupled to a radiation field, it is one of the paradigms of quantum optics.
Another paradigm is stimulated emission, which leads for the case of a large number of atoms to the
concept of a laser. The corresponding concept in the case of spontaneous emission of an ensemble with a large 
number of atoms is the {\em superradiator}. This system has been first proposed by Dicke in 1954 \cite{Dic54}, but it took
nearly 20 years for the first experiment \cite{Skretal73} to be carried out to observe the predicted superradiant emission peak
with an intensity proportional to the {\em square} of the number of atoms (molecules) of the emitting gas. Still, this
effect has never become as popular as the laser, not to speak of a use in the form of a device. Nevertheless, the physics
of superradiance is one of the most interesting in the field of quantum physics, because it comprises a number of
fundamental concepts such as coherence, symmetry, interaction between particles,  the non-equilibrium physics of
transient processes, and the notion of a quasiclassical limit \cite{Andreev,Benedict}.

In this short paper, we discuss the superradiance effect in a different context, that is in solid state physics.
In a recent calculation \cite{BIS98}, we predicted a novel form of superradiance for an open system, i.e. an optical active region that
is pumped externally
by electron (hole) reservoirs. In semiconductor quantum wells, the coherent decay of electron-hole pairs leads to
a peak of the emitted light  with a strong intensity that, as a function of time, shows oscillations with a frequency
\begin{equation}
\label{frequency}
\omega \simeq \sqrt{2\Gamma T},
\end{equation}
where $\Gamma $ is the spontaneous decay rate of a single pair and $T$ the rate at which electrons
are pumped in the conduction and holes into the valence band. The system is under a strong magnetic field which guarantees
that all optical matrix elements for the recombination of individual electron-hole pairs are the same.


\section{Model and results}

The photon field 
$H_{p}= \sum_{{\bf Q}}\Omega_{{\bf Q}}a^{\dagger}_{{\bf Q}}a_{{\bf Q}}$
with creation 
operator $a^{\dagger}_{{\bf Q}}$ for a mode ${\bf Q}$ gives rise to 
transitions which change  an internal degree of freedom $\sigma=(\uparrow, 
\downarrow) $ of one-particle electronic states labeled $(i,\sigma )$ with creation
operator $c^{\dagger}_{i,\sigma}$. The states 
are degenerate with respect to $i$ with energies $\varepsilon_{i 
\uparrow}=-\varepsilon _{i\downarrow}=\hbar\omega _{0}/2$. The 
electron--photon coupling matrix element $g_{{\bf Q}}$ is assumed to be 
independent of the electronic quantum numbers $(i,\sigma )$. 
Then, the Hamiltonian of the optically active region
can be written as 
\begin{equation}
\label{dhamiltonian}
H_{D}=\hbar\omega _{0}\hat{J}_{z}+
\sum_{{\bf Q}}
g_{{\bf Q}}\left(a^{\dagger}_{{\bf Q}}+a_{{\bf Q}}\right) 
\left(\hat{J}_{+}+\hat{J}_{-}\right)+ H_{p},
\end{equation}
where the operators   
$\hat{J}_+:=\sum_i c^{\dagger}_{i,\uparrow }c_{i,\downarrow }$,
$\hat{J}_-:=\sum_i c^{\dagger}_{i,\downarrow }c_{i,\uparrow }$
and
$\hat{J}_z:=\frac{1}{2}\sum_i \left(c^{\dagger}_{i,\uparrow }c_{i,\uparrow }
-c^{\dagger}_{i,\downarrow }c_{i,\downarrow }\right)$
form a (pseudo) spin algebra with angular momentum commutation relations.

We allow the number of electrons $N$ in the active region 
to vary by tunneling to and from electron reservoirs $\alpha= I/O$ 
(`In'and `Out') with Hamiltonians 
$
H_{\alpha }=\sum_{k}\varepsilon _{k}^{\alpha }
c^{\dagger}_{k,\alpha }c_{k,\alpha }
$
for non-interacting electrons 
with equilibrium Fermi distributions $f_{\alpha 
}$.
The tunneling of electrons is described by the usual tunnel Hamiltonian
$
H_{T}=\sum_{ki\sigma\alpha}\left( t^{\alpha }_{ki\sigma } 
c^{\dagger}_{k,\alpha }c_{i,\sigma  } + c.c. \right)
$
with coefficients $t^{\alpha }_{ki\sigma }$. 
The total Hamiltonian is given by the sum $H=H_{D}+\sum_{\alpha}
H_{\alpha }+H_{T}$. 

The active region is characterized by so-called Dicke eigenstates of the total pseudo 
spin $J$ and its projection $M$ through  
$\hat{J}^{2}  | JM;\{\lambda\} 
\rangle=  J(J+1)   | JM;\{\lambda\} \rangle$ 
and $\hat{J}_z  | JM;\{\lambda\} 
\rangle=  M   | JM;\{\lambda\} \rangle$, 
where $\hat{J}$ is the total pseudo 
spin operator. Here, $\{\lambda\} $ denotes all additional quantum numbers 
apart from $J$ and $M$ that are necessary to characterize the eigenstates of 
$H_D$. Radiative transitions obey the selection rule $M\to M\pm 1$ with a 
spontaneous emission intensity $I_{JM}= \hbar 
\omega_{0} \Gamma \nu_{JM}$, $\nu_{JM}:= (J+M) (J-M+1)$. 
Here, $\Gamma$ is the spontaneous emission rate of 
one {\em single} two-level 
system; for radiative transition in atoms, $1/\Gamma $ is in the nano second 
range.

In order to describe the time evolution of the active region coupled to the reservoirs,
we used a master equation description in the basis of the Dicke eigenstates. The only relevant quantum
numbers in this effective description are $J$, the length of the pseudo-spin, and $M$, its projection.
We chose boundary conditions in analogy to electronic pumping in semiconductor laser diodes: each electron
entering or leaving the active region leads to an increase of $M$. On the other hand, the collective
radiative decay decreases M. The dynamics of the pseudo spin therefore is driven by the combined influence
of two `forces'. In fact, in the classical limit of the master equation, we find the dynamics described by
an equation in the $J$-$M$ phase space,
\begin{eqnarray}
\label{eom}
\dot{M}(t)&=&-\Gamma\nu_{J(t)M(t)}+T\nonumber\\ 
\dot{J}(t)&=& T \cdot M(t) /J(t).
\end{eqnarray} 
For $T>2\Gamma$, 
this describes a harmonic oscillator  
with frequency Eq.~(\ref{frequency}) and amplitude dependent 
damping. For $T\to 0$, there is a smooth 
crossover to the conventional Dicke peak with vanishing intensity at large 
times and without oscillations.

\section{Discussion and outlook}

Our model applies to the small sample limit of the superradiant problem:
reabsorption processes of photons that may lead to oscillatory behavior of 
the intensity do not play any role here. In fact, that kind of
oscillations in  superradiance has already been observed in the first experimental
verification of the Dicke effect by Skribanowitz et al. \cite{Skretal73}. They are due to a flow
of energy between the atoms and the field modes and can be considered as generalized
Rabi-oscillations. Their frequency $\omega$ can be estimated by solving for a mean field theory
of the Maxwell-Bloch equations in a one-dimensional model \cite{Benedict} with the result
$\omega\sim N^{1/2} d/\ln N$, where $d$ is the modulus of the dipol matrix element. 

In our model, the oscillations are due to the combination of two mechanisms (decay and pumping),
where the backflow of energy from the field is irrelevant. The dependence of the frequency
$\omega \simeq \sqrt{2\Gamma T}$ on the pumping (transmission $T$) should be used to identify
this kind of oscillation, which is similar to
relaxation oscillations in semiconductor laser diodes. We point our here, however, that the physics
of superradiance is different from that of a laser because not stimulated but spontaneous emission is
the basic mechanism.

To observe this kind of superradiance experimentally, we propose the system of electrons and holes in
a semiconductor quantum well in a strong magnetic field. Electrons are injected into the conduction band and
holes into the valence band of the active region, either by vertical tunneling or thermal emission.
The strong magnetic field is necessary to have dispersionless 
single electron 
levels $i=X$, corresponding to the lowest 
Landau bands ($n=0$) and guiding 
center $X$ in the conduction and the valence bands. 
In this case, the interband optical 
matrix elements are diagonal in $i$.

Superradiance in {\em bulk} semiconductors under strong magnetic fields has already been discussed theoretically
in the literature \cite{BKK}. The recombination of free electrons and holes was found to yield a short and powerful
pulse of coherent light of duration 0.1-1 ps and intensities of larger than 100 MW cm$^{-2}$, with stationary
or pulsed magnetic fields of larger than $10^5$ G.

Our proposal is aimed at a situation which is closer to quantum-Hall conditions, where the kinetic energy of
the electrons is quenched in two dimensions. Furthermore, our calculations support that the initial
peak of superradiance is strongly enhanced by the electronic pumping so that in principle one should
arrive at very short light pulses of high intensity. The time scale to observe this effect is limited by
electron-electron scattering which, as a dephasing mechanism, has to be reduced in order to obtain the collective
build-up of the large pseudo spin that decays coherently. Typical temperatures 
should then be in the sub-Kelvin regime, with magnetic fields of a few Tesla. The subsequent oscillations should also be
observable in corresponding oscillations of the pumping current around its stationary value. A preliminary
estimate of the time scales then yields an oscillation period of 1 ps for a 1 mA current, with pulse durations
shorter than 1 ps and a dephasing time of $ 10^{-11} s$.

\end{document}